# Towards the cycle structures in complex network: A new perspective

Tianlong Fan[1,2], Linyuan Lü[1,2*] & Dinghua Shi[3*]

**Stars and cycles are basic structures in network construction. The former has been well studied in network analysis, while the latter attracted rare attention. A node together with its neighbors constitute a neighborhood star-structure where the basic assumption is two nodes interact through their direct connection. A cycle is a closed loop with many nodes who can influence each other even without direct connection. Here we show their difference and relationship in understanding network structure and function. We define two cycle-based node characteristics, namely cycle number and cycle ratio, which can be used to measure a node's importance. Numerical analyses on six disparate real networks suggest that the nodes with higher cycle ratio are more important to network connectivity, while cycle number can better quantify a node influence of cycle-based spreading than the common star-based node centralities. We also find that an ordinary network can be converted into a hypernetwork by considering its basic cycles as hyperedges, meanwhile, a new matrix called the cycle number matrix is captured. We hope that this paper can open a new direction of understanding both local and global structures of network and its function.**

The last two decades have witnessed a great development in network science studies. The three classical network models are random networks[1,2], small-world networks[3], and scale-free networks[4], based on which the researchers have carried out a wealth of experimental or theoretical analysis on both real and modeled networks. Most of the studied networks, whether naturally formed such as neural networks, food webs and gene-protein regulatory networks, or constructed by human society like transportation networks, power networks and social networks, are considered based on the chain-structure and neighborhood star-structure. In fact, beyond the chains and stars, a complex network also has abundant cycle-structures[5], see examples in Figure 1a, 1b and 1c. As far as interaction be concerned, the star-structure assumes that the interactions between two nodes can only be achieved through their direct connections, but for cycle-structure, a node can influence others on the same cycle even they are not directly connected. The difference of these two viewpoints, i.e. star- and cycle-structure, is just like the vivid description in the poem

[1] Institute of Fundamental and Frontier Sciences, University of Electronic Science and Technology of China, Chengdu 611731, China. [2] Alibaba Research Center for Complexity Sciences, Alibaba Business College, Hangzhou Normal University, Hangzhou 311121, China. [3] Department of Mathematics, Shanghai University, Shanghai 200444, China. *Correspondence and requests for materials should be addressed to L. L. (email: linyuan.lv@gmail.com) or to D. S. (email: shidh2012@sina.com).

*Written on the Wall at West Forest Temple* written by Su Tung-P'o, a great Chinese poet of the Song Dynasty, which reads:

> From the side, a whole range; from the end, a single peak:
>
> Far, near, high, low, no two parts alike.
>
> Why can't I tell the true shape of Lu-shan?
>
> Because I myself am in the mountain.

(Translated by American scholar Burton Watson[6]).

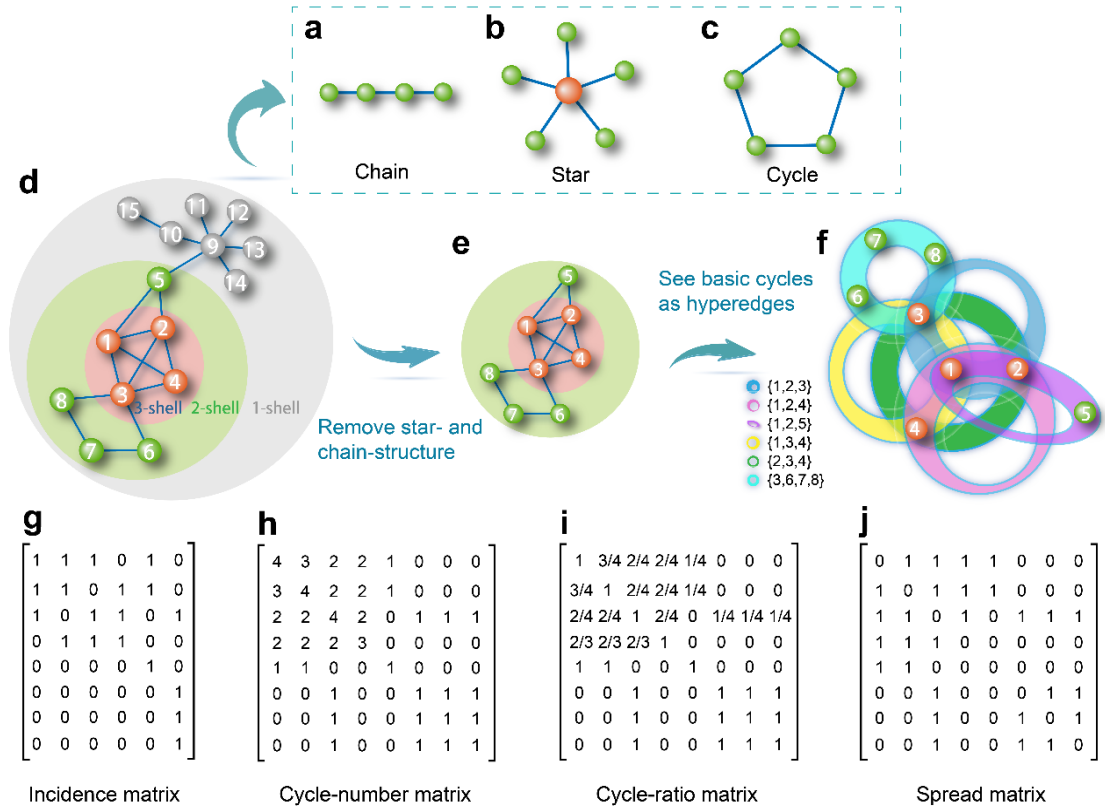

**Figure 1. The three types of basic structures, transformation and four cycle-based matrices of an exampled network.** Here **a**, **b** and **c** are three basic structures, chain, star and cycle respectively, of the example network **d** which has three layers according to the network *K*-core decomposition. **e** is a sub-network of **d**, obtained by removing the chain- and star- structure in the periphery. **f** is the hypernetwork converted from **e** by considering all six basic cycles as hyperedges. Each ring represents a hyperedge. **g** is the incidence matrix of **f**, in which rows represent nodes and columns represent hyperedges. **h** is the cycle number matrix of **e** and **f**, which can be calculated by multiplying the incidence matrix **g** by its transposed matrix. **i** is the cycle ratio matrix which is obtained by dividing each of the elements in the cycle number matrix by the diagonal elements, and the sum of the elements in each column equals to the cycle ratio of the corresponding node that this column represents. **j** is the corresponding spread matrix which is obtained by firstly mapping the cycle number matrix to a binary matrix and then setting the diagonal elements 0.

Chains, stars and cycles are the basic structures of a network and have significant roles in network construction. Three typical examples are small-world network, scale-free network and totally homogeneous network, respectively. A small-world network is constructed by randomly rewiring links of a regular network, the rewiring process is actually an operation based on chain structure, which has a great impact on the network structure and function[7]. By rewiring the links, a few small cycles are broken maintaining the relative large average clustering coefficient[3], while the addition of shortcuts shorten the average shortest path of the network. The scale-free network model, i.e., the BA model, based on the growth and preferential attachment rules[4] reveals the formation mechanism of the "Matthew effect". In the model, each step a new node with a few links are added to the network, implying that the BA model is actually a growing model based on star-structure (considering the new added node as the center node of the star). To achieve better synchronization, Shi et al.[5] proposed a totally homogeneous network where the network nodes have the same degree, node girth (i.e., edge number of the node' smallest cycle), and node path-sum (i.e., total number of edges from other nodes to node $i$ through shortest paths). It was found that the longer the girth and the shorter the path-sum of the network, the easier it is to synchronize, reflecting the important role of cycle-structures in network synchronization. Subsequently, Sizemore et al.[8] identified two particular substructures in human brain network, namely cliques and cavities, which are crucial for human cognitive processes and complex behaviors. In fact, cliques and smallest cavities are essentially totally homogeneous networks or the sub-networks.

In addition to the model world, in reality, examples that involve cycle-structure are ubiquitous in various fields, such as a mass of communication ways, Internet applications and social organization, which are more about the interaction among three or more subjects. As opposed to point-to-point communication, meeting is the most traditional way of human communication based on cycle-structure which has expanded to online society with the development of web technologies. Comparing with the offline activities, online communications are more convenient, efficient and frequent. Typical examples of cycle-based communication include online forums, Facebook's 'Event', Weibo's 'Topic', and the 'Group chats' of social media like WeChat and QQ. To analyze such systems from the perspective of cycle-structure, it requires new network science theories and tools. Although Berge[9] has proposed hypergraph theory in 1970s to describe the interactions among at least three nodes, it still lacks a proper mathematical representation. Therefore, over the past forty years, it has been difficult for researchers to advance hyper-network-related researches. Cycle-based communication is actually a change in the mode of information dissemination, which will further affect the organization of human social activities. The trend of 'Internet of Everything' will promote the development of cycle-based interaction mode in the field of technology and production.

From ancient Greece to today, the cycle has always been regarded as the most perfect geometric shape. As a basic structure, it is filled with all aspects of our life with its beautiful shape and excellent symmetry, balance and feedback features. Recent studies have shown that the cycle-like neural paths are important for human cognitive processes and complex behaviors[8], and thus very crucial structure in the brain. Lou et al.[10] found that the $q$-snapback network composed of many chain- and loop-motifs has the strongest robustness of controllability. Building new network configuration model based on the distribution of shortest cycle lengths has become a new direction of random network research[11]. In the research of RNAs and various diseases related to gene expression, circular RNAs (or circRNAs, a class of non-coding RNAs that may regulate gene expression in mammals, circRNAs are characterized by covalently closed loop structures.) has become a new hotspot in recent years due to its greater stability, universality, conservation and tissue specificity. Han *et al.*[12] identified a kind of circRNA that can significantly down-regulate in HCC (human hepatocellular carcinoma) tissues and closely related to the prognosis of HCC patients, therefore this circRNA is expected to be a potential therapeutic target for HCC. Social networking is also related to cycle. Two persons are more likely to be friend if they have more common friends. Based on this fact, a recommendation algorithm called Common Neighbor[13] was proposed to recommend new friend. Once new relationship exist, new cycles (i.e., triangles or higher-order cycles, like *k*-clique) are formed.

Many real networks have rich cycle-structures but were studied mostly from star-structure perspective. For example, coreness denoted by $k_S$ is a widely used node centrality for identifying influential nodes on networks. It can be calculated by *K*-core decomposition, a method from star perspective (see Ref. 14 for detailed process), which divides the network into many layers. Figure 1d shows an example network with three layers. A node with larger coreness value is more centrally located in the network, and thus has higher influence than nodes located in the periphery[14] (usually leaves or chains). For *K*-core decomposition, it has long been neglected that the nodes with $k_S=1$ must be chain-structures if they are not neighborhood star-structure; and the nodes with $k_S \geqslant 2$ must be in at least one cycle. In fact, in addition to layering the nodes, the *K*-core decomposition also layers the global network structures. For example, the three layers in Figure 1d correspond to star and chain (periphery), cycle (middle) and clique (center) structures, respectively. In terms of connectivity, cycle is an important topological object between the two extremes of star- and clique-structures. Lü *et al.*[15] showed the graceful relationship among degree, *H*-index and coreness of network nodes, called the DHC Theorem on networks. This relation provides a decentralized way to calculate coreness, and obtains a family of *H*-indices which are essentially analogical with the only difference lie in the degree to which the star-structures are measured. Compared with the microscopic and centralized star-structure, the cycle-structure is a mesoscale and decentralized network architecture, and thus an important complement of the star-structure.

Based on the view of cycle-structure, this paper proposes two new indicators for identifying important node: cycle number and cycle ratio. Considering information propagation on cycle, a spread matrix is defined as a mathematical representation tool for new information dissemination model. The simulation results on real networks show that in terms of network connectivity, the networks collapse faster when removing nodes according to the descending order of their cycle ratio values. On the other hand, for network information dissemination, more nodes are infected when propagating initialization priority is given to the nodes with higher values of cycle number.

Furthermore, if we consider each simple cycle in an ordinary network as a hyperedge, the ordinary network can easily be converted into a hypernetwork, see an example in Figure 1f. Then the ordinary network can be described by the incidence matrix of its corresponding hypernetwork. The incidence matrix multiplied by its transposed matrix obtain the cycle number matrix, which is closely related to the spread matrix of the ordinary network. Conversely, we may also think of a hyperedge as a basic cycle, and thereby obtain a new perspective for insight into hypernetworks. The conversion relationship between the hypernetwork and its corresponding ordinary network, as well as the incidence matrix, the cycle number matrix, and the spread matrix, provide useful tools for analyzing the structure and dynamics on hypernetworks mathematically.

In this paper, we mainly discuss the cycle in the usual sense, namely 1-cycle, which is a closed planar structure surrounded by nodes and edges. However 2-cycle, closed structures enclosed by triangles, and other higher-order cycles are also rich in complex networks. It is alternative to use algebraic topology[16] tools for reference to understand and study higher-order cycles which is not as intuitive as 1-cycle.

## Results

We firstly introduce some definitions related to cycle-structure.

**Cycle**: a path $P = v_1v_2…v_k$ is a cycle if it satisfies the three conditions: (i) $k > 2$; (ii) the previous $k-1$ nodes are different from each other; (iii) $v_1 = v_k$.

**Smallest cycle of a node**: cycles of node $i$ with the least number of edges are called the smallest cycles of node $i$, and the number of edges is called its girth[5].

**Redundant cycle** and **basic cycle**: a cycle is redundant, if it contains smaller cycles, otherwise it is called a basic cycle. The exact definition requires algebraic topological tools[16].

**Smallest basic cycle of a network**: the set consisted of the smallest cycles of all nodes is called the smallest basic cycles of network.

**Smallest basic cycle of a node**: the network smallest basic cycles that contain node $i$ are called the smallest basic cycles of node $i$.

**Clique** and **cavity**[8]: a cycle is called a clique if it is fully connected, otherwise it is called a cavity which requires algebraic topological tools to give its exact definition[16]. We denote a clique as $k$-clique if it has $k+1$ nodes.

Take Figure 1e as an example, the different types of cycles defined above are shown in Table 1. Note that, in this example, the set of basic cycles is exactly the same as the set of its smallest basic cycles. However, in most cases, they are usually different.

**Table 1 | The cycles of different types in Figure 1e**

| | | |
|---|---|---|
| **Node 3's** | Smallest cycles | {3, 1, 2}, {3, 1, 4}, {3, 2, 4} |
| | Basic cycles | {3, 1, 2}, {3, 1, 4}, {3, 2, 4}, {3, 6, 7, 8} |
| | Redundant cycles | {3, 1, 2, 4}, {3, 1, 5, 2}, {3, 1, 5, 2, 4} |
| **Network's** | Basic cycles | {1, 2, 3}, {1, 2, 4}, {1, 2, 5}, {1, 3, 4}, {2, 3, 4}, {3, 6, 7, 8} |
| | Cavity | {3, 6, 7, 8} |
| | 2-Clique | {1, 2, 3}, {1, 2, 4}, {1, 2, 5}, {1, 3, 4}, {2, 3, 4} |
| | 3-Clique | {1, 2, 3, 4} |

Based on the above concepts, we define two nodes importance ranking indicators.

**Cycle number**: the number of basic cycles through node $i$ is called cycle number of node $i$.

In real network analysis, searching all basic cycles is very time consuming especially for large networks. Therefore, when using the cycle number to evaluate the node's importance, we may have to make a trade-off between accuracy and efficiency. An alternative is only considering the smallest basic cycles of node, see a search algorithm for basic cycles in *Supplementary Information* section 1.

**Cycle ratio**: the sum of the proportions of the node $i$ appearing in the basic cycles of the nodes contained in basic cycles of node $i$. Taking node 1 in Figure 1e as an example, all the basic cycles of node 1 involve a total of five nodes, namely nodes 1-5. For each node among the five, calculate the proportion of node 1 appearing in its basic cycles, and finally summarize all five ratios to obtain the cycle ratio of node 1: 4/4 + 3/4 + 2/4 + 2/3 + 1/1 = 47/12. Table 2 shows the basic cycles, cycle number and cycle ratio of the nodes in Figure 1e.

**Table 2 | Basic cycles and two cycle-based importance indicators of the nodes in Figure 1e.**

| Node Label | Basic cycles | Cycle number | Cycle ratio |
|---|---|---|---|
| 1 | {1, 2, 3}, {1, 2, 4}, {1, 4, 3}, {1, 5, 2} | 4 | 47/12 |
| 2 | {2, 3, 1}, {2, 4, 1}, {2, 1, 5}, {2, 4, 3} | 4 | 47/12 |
| 3 | {3, 1, 2}, {3, 1, 4}, {3, 2, 4}, {3, 6, 7, 8} | 4 | 68/12 |
| 4 | {4, 1, 2}, {4, 1, 3}, {4, 2, 3} | 3 | 10/4 |
| 5 | {5, 2, 1} | 1 | 6/4 |
| 6 | {6, 7, 8, 3} | 1 | 13/4 |
| 7 | {7, 8, 3, 6} | 1 | 13/4 |
| 8 | {8, 3, 6, 7} | 1 | 13/4 |

Note that the basic cycles of the nodes are exactly the same as their smallest basic cycles.

**Hypergraph**: let $H = (V, E)$ be a hypergraph[9], where $V = \{v_1, v_2, ..., v_i, ...\}$ ($1 \leq i \leq |V|$) and

$E = \{e_1, e_2, \ldots, e_j, \ldots\}$ $(1 \leq j \leq |E|)$ are the sets of nodes and hyperedges, respectively. The $|V|$ and $|E|$ are number of nodes and hyperedges, respectively.

**Incidence matrix**: for a hypergraph $H = (V, E)$, its incidence matrix is denoted by a $|V| \times |E|$ matrix $M(H) = [h_{ij}]_{|V| \times |E|}$ where the element $h_{ij} = 1$ when the nodes $v_i$ is in the hyperedge $e_i$, otherwise $h_{ij} = 0$. In the incidence matrix, each row represents a node in $V$ and each column represents a hyperedge in $E$, see Figure 1g the incidence matrix of network in Figure 1f.

**Cycle number matrix**: given an ordinary network $G(V, E)$ with cycle-structures, its cycle number matrix is denoted by a matrix $B(G) = [b_{ij}]_{|V| \times |V|}$ where the element $b_{ij}$ is the number of co-cycles of node $i$ and $j$ when $i \neq j$, and the cycle number of node $i$ when $i = j$. If the basic cycles of $G(V, E)$ are regarded as hyperedges, then the cycle number matrix $B(G)$ can be obtained by multiplying the incidence matrix of the hypernetwork $M(H)$ by its transposed matrix $M^T(H)$, namely $B(G) = M(H) \times M^T(H)$, see Figure 1h the cycle number matrix of network in Figure 1e.

On the other hand, for $G(V, E)$, if all the elements in its cycle number matrix are divided by their corresponding diagonal elements of the respective rows, then we obtain the cycle ratio matrix, see an example in Figure 1i, and the sum of the elements in each column equals to the cycle ratio of the corresponding node that this column represent.

**Spread matrix**: given network $G(V, E)$ the spread matrix denote as $S(G)$ is obtained by mapping the $B(G)$ into a binary matrix where all the diagonal elements are 0, and the other elements are 1 if their corresponding values in $B(G)$ are greater than 1, otherwise 0, see Figure 1j the spread matrix of network in Figure 1e. In fact, the spread matrix can be defined in different ways according to the specific research contents, other two examples are shown in *Methods* and *Supplementary Information* section 2. Similar to the significance of adjacency matrix for the network study from star-structure perspective, the spread matrix can be used to study the structures and dynamics on hypernetworks and ordinary networks from cycle-structure perspective.

In order to analyze the performance of new proposed node importance ranking indicators based on cycle-structure, we consider its importance to network connectivity and propagation capability, and compare them with some representative ranking measures based on star-structure. On the one hand, we simulate node-removal attacks according to the descending order of the values given by different indicators, and observe which indicator can dismantle the network faster. On the other hand, we perform the propagation experiments according to the cycle-based SIR model. For each node, we set it as the initial seed and it will spread the information according to the spread matrix. To avoid the fluctuations in single realization, we use the average number of final affected nodes of all realizations to characterize the node propagation capability. Finally, we calculate the Kendall Tau correlation coefficient[17] between the ranking list given by each indicator and the corresponding node propagation capability list. The coefficient is in the range [-1, 1], the larger the $\tau$, the better performance of the indicator on quantifying the node propagation capability.

The experiments were carried out on six real undirected and unweighted networks from diverse fields, including the neural network of the nematode C. elegans[18] (C. elegans), the Email network[19] (Email), the trans-European road network[20] (EuroRoad), the co-authorship network of scientists working on network science[21] (NS-GC), the US Air lines network[15] (USAir), and the protein-protein interaction network in yeast[22] (Yeast). Their basic topological features are shown in Table 3. Besides, the probability distribution of basic cycles of all different lengths are compared between real networks and model networks (including small-world network, scale-free network and random network), and significant differences were found, see *Supplementary Information* section 3, indicating that these models cannot well quantify the mesostructure features of real-world networks in the view of cycles.

**Table 3 | The basic topological features of the six real networks**

| Networks  | $|V|$ | $|E|$ | $\langle k \rangle$ | $\langle d \rangle$ | $C$   |
|-----------|-------|-------|---------------------|---------------------|-------|
| C. elegans| 306   | 2148  | 7.89                | 2.45                | 0.31  |
| Email     | 1133  | 5451  | 9.62                | 3.61                | 0.25  |
| EuroRoad  | 1174  | 1417  | 1.21                | 18.37               | 0.02  |
| NS-GC     | 379   | 914   | 4.82                | 6.04                | 0.80  |
| USAir     | 332   | 2126  | 12.81               | 2.74                | 0.749 |
| Yeast     | 2375  | 11693 | 4.92                | 5.09                | 0.388 |

Here $|V|$ and $|E|$ are the number of nodes and edges, respectively, $<k>$ and $<d>$ are the average degree and the average shortest distance, respectively, and $C$ is the clustering coefficient.

We perform node-removal attacks according to five node importance indicators: degree, *H*-index, coreness, cycle number and cycle ratio. Two cyclic indicators are based on the node's smallest basic cycles. For each indicator, nodes with higher values will be removed first from the network. The dependence of the proportion of the largest connected component on the ratio of removed nodes is shown in Figure 2. Generally speaking, the network collapses faster when attacked by cycle ratio comparing with other indicators, and this advantage keeps stable on all the investigated networks.

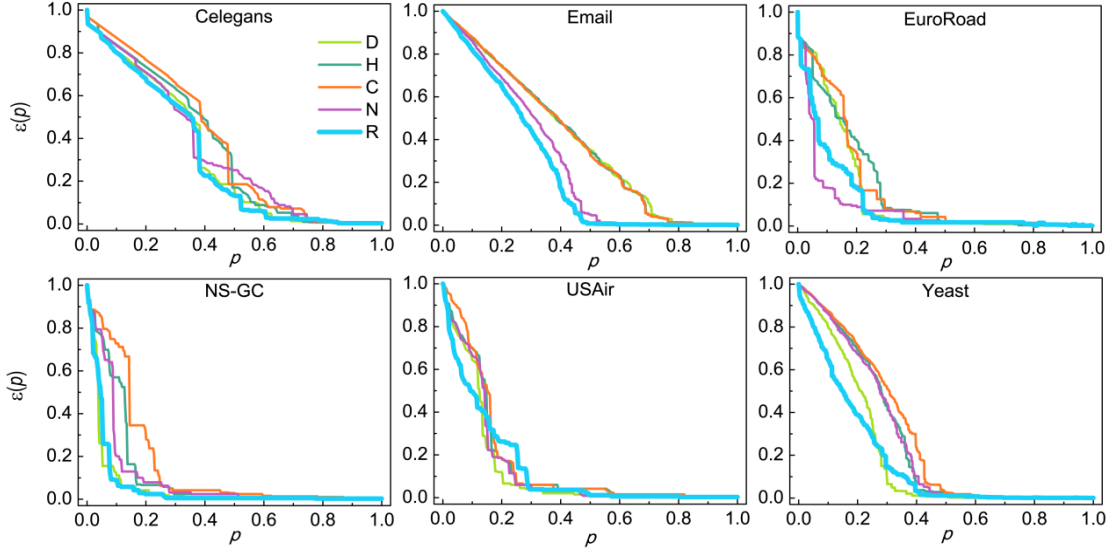

**Figure 2. The results of node-removal attacks on six networks.** The $p$ is the ratio of removed nodes which are ranked in descending order according to degree (D), $H$-index (H), coreness (C), cycle number (N), and cycle ratio (R), respectively. The $\varepsilon(p)$ is the proportion of the largest connected component in the remaining network after removing nodes of fraction $p$.

Propagation based on star-structure is along the links between two nodes, which has been well studied in literatures. However, the cycle-based propagation has attracted less attention and is rarely studied. Unlike the star-based propagation, cycle-based propagation allows the nodes who are together in a smallest basic cycle directly influence each other even they are not directly connected. As an example, we refer to the classic SIR (susceptible-infected-recovered)[23,24] propagation model which is based on the star-structure, and propose a new cycle-based SIR model, denote by cSIR, where the information can be propagated along cycles. That is to say, if the node $i$ is infected, then all the susceptible nodes on its smallest basic cycles are at risk of being infected, regardless of whether there is a direct connection between the susceptible node and node $i$.

If we call the original SIR model a point-to-point propagation pattern, then the cSIR model figures a point-to-surface pattern which is obviously more efficient for propagation. Represented by online social platforms such as Facebook and Weibo, the information dissemination way based on cycles is penetrating into many areas of our life with the help of modern communication technology.

We compare the results of propagation on six real networks based on the original SIR model and the cSIR model where the smallest basic cycles are considered. For the detailed processes of these two propagation model can be found in *Methods*. For each network, we select the node with the maximum cycle number as initial seed for propagation, and the change of the number of infected nodes over time is shown in Figure 3, two cycle indicators are based on the node's smallest basic cycles and the results are averaged over 100 realizations. To make fairly comparison, the propagation probability of cSIR model are set to be the same as the original SIR model, see Table S1 in *Supplementary Information* section 2, namely $\beta$=0.09 for C. elegans,

$β$=0.14 for Email, $β$=0.50 for EuroRoad, $β$=0.36 for NS-GC, $β$=0.06 for USAir, and $β = 0.15$ for Yeast, respectively.

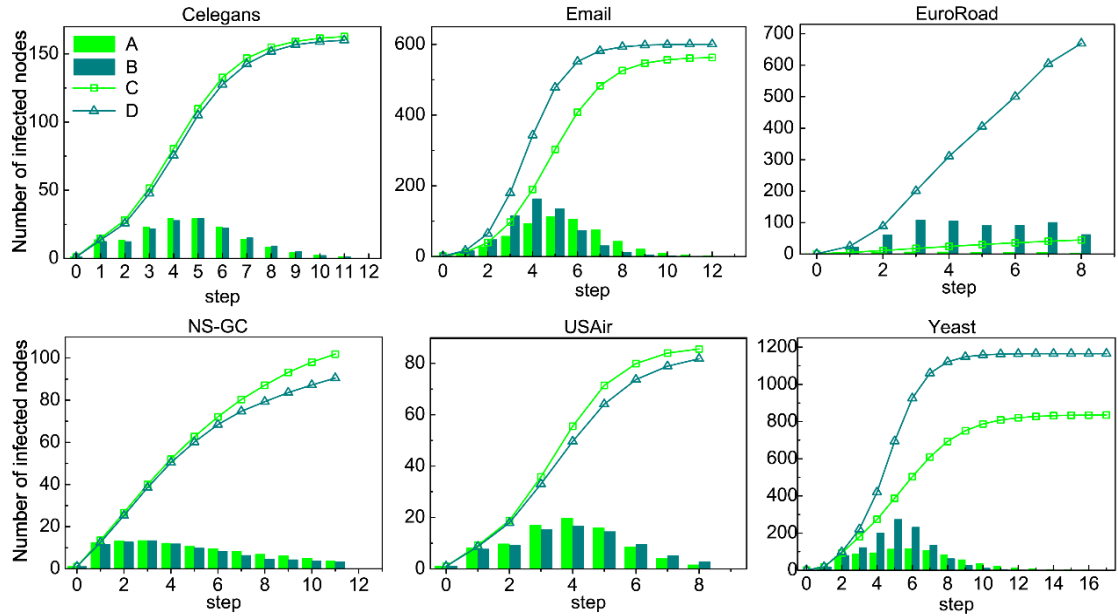

**Figure 3. Comparison of the propagation speed and capacity between the original SIR model and the cSIR model.** The green and blue histograms (A and B) are the number of new infected nodes in each step of SIR and cSIR models, respectively, while the green and blue lines (C and D) represent the cumulative number of infected nodes of SIR and cSIR models, respectively. All results are averaged over 100 realizations.

Table 4 is the Kendall Tau correlation coefficient between ranking lists given by the five indicators and the cSIR model. Two cycle indicators are based on the node's smallest basic cycles. Here we simulated all $β$ values in the range [0, 1] in increments of 0.01, and then selected the overall best groups. In four of the six networks, the cycle number performs the best, while for the rest two networks, it is near the best.

**Table 4 | The Kendall Tau between the ranking list given by each indicator and the corresponding node propagation capability given by the cSIR model.**

| Networks | Degree | $H$-index | Coreness | Cycle ratio | Cycle number |
|---|---|---|---|---|---|
| C. elegans ($β = 0.14$) | 0.8184 | 0.8299 | 0.7805 | 0.6540 | **0.8339** |
| Email ($β = 0.22$) | 0.7358 | 0.7296 | 0.7110 | 0.6995 | **0.8033** |
| EuroRoad ($β = 0.02$) | 0.5091 | 0.5719 | 0.7282 | **0.8319** | 0.7789 |
| NS-GC ($β = 0.06$） | 0.7168 | 0.7248 | 0.6851 | 0.4224 | **0.7269** |
| USAir ($β = 0.14$） | 0.8767 | **0.9075** | 0.9072 | 0.4951 | 0.8848 |
| Yeast ($β = 0.14$） | 0.7209 | 0.7173 | 0.7013 | 0.6058 | **0.7854** |

The $β$ is the propagation probability. In each row, the largest $τ$ is highlighted in bold.

From above comprehensive analysis, we can conclude that, in terms of the node importance to network structure, taking connectivity as an example, the nodes-removal attack is more effective according to the cycle ratio; in terms of the node importance to network function, taking the propagation capability as an example, the nodes with larger cycle number can affect more nodes during the propagation process.

**Discussion**

We proposed a new perspective of cycles to analyze the networks. A cycle is a closed loop formed by at least three nodes. It exists in a large number of various real-world and model networks. As a mesostructure in the networks, cycles play important role in network structure and function, but have been neglected in the network analysis and modeling. Comparisons between the real networks and modeled networks, including the WS model (small-world networks) and the BA model (scale-free networks), showed the large difference on the distribution of basic cycles, indicating the weakness of the models for studying the real networks. Therefore, it is worthwhile to systematically study the cycle-structure in network.

To quantifying the node's importance, two cycle-based indicators were defined, namely cycle number and cycle ratio. The cycle number of a node describes the degree of local clustering around the node, and the cycle ratio delineates how tightly a node is connected to all of the nodes on its basic cycles. Experiments on six real networks showed that, the nodes with higher cycle ratio are more importance to the network connectivity, while the nodes with higher cycle number have higher propagation capability under the cycle-based SIR model.

Based on the cycle-structure, some useful tools were provided for mathematically analyzing the structure and dynamics on hypernetworks. From cycle perspective, an ordinary network can be converted into a hypernetwork by considering its basic cycles as hyperedges. Now the questions are whether the reverse can be performed, and if so, whether the reverse process preserves all information. Since ordinary networks are special cases of hypernetworks, it is clear that generally the answers are no. However, this does not exclude some particular situations. Thus, the new notion sheds some light on future research on hypernetworks.

Similar to the analysis on nodes, one can also define the measures for edges, such as the cycle number and cycle ratio of edges that can be used to quantifying the edge importance. Besides, the cycle-based network analysis may shed some light on the design of link prediction algorithms[25], such as the well-known Common Neighbor method which considers the number of new added 2-clique after the addition of a link as proportional to its existence probability, and the loop model[26].

We believe that the cycle-structure is an important complement and breakthrough of the existing star-structure. This paper, from cycle perspective, is expected to open up a new direction for the study of the mesoscale and global structure of the networks. As a starting point, the paper only considered the 1-cycle analysis, however, the higher-order cycles, see examples in Figure 4, are also interesting and worth studying, which will be our future research. The algebraic topology[16,27] will become a powerful tool for the high-order cycles related studies.

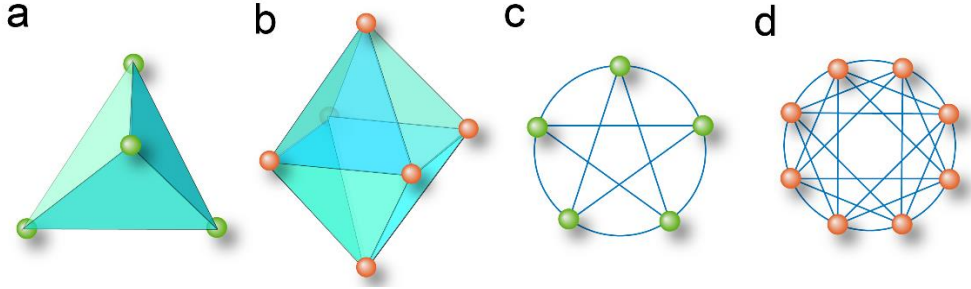

**Figure 4. Example of higher-order cycles.** Here **a** and **b** are 2-cycles: 3-clique and the smallest 2-cavity, respectively[8]. **c** and **d** are 3-cycles that are all surrounded by tetrahedrons: 4-clique and the smallest 3-cavity, respectively. They are all totally homogeneous networks.

## Methods

**Propagation models**

**1. The standard SIR model** The susceptible-infected-recovered (SIR) epidemic spreading model was usually applied to study the spread of disease or messages on networks. First the target nodes are initialized to the infected state, and the other nodes are susceptible. At each time step, each infected node will infect all its neighbor nodes in the susceptible state through the direct connections with probability $\beta$, and then becomes the recovered state with probability $\lambda$ (here we set $\lambda = 1$). This propagation process is repeated until there is no node in the infected state in network. To quantify the propagating capacity of a node, we initialize only one node into the infected state, and the cumulative number of final infected nodes trigger by this node will be calculated to quantify its propagating capacity. The more nodes infected by a node, the higher the propagating capacity of this node. We take the propagation threshold, $\beta_c \approx \langle k \rangle / (\langle k^2 \rangle - \langle k \rangle)$, according to the heterogeneous mean-field theory[24] as a reference, and simulate with different propagation probability[28,29], including $\beta = \beta_c$, $2.5\beta_c$, and $5\beta_c$. Finally, we take the average value of 100 independent experiments to eliminate the fluctuations of the simulations.

**2. The cycle-based SIR (cSIR) model** The cycle-based SIR model is different from the standard SIR model because a node can infect all other nodes on its basic cycle simultaneously, even though there are not a direct connection between them. Conversely, if a node and one of its neighbors do not form a basic cycle, then the information cannot be propagated through the direct connection between them. At each step, the nodes in the infected state infect all of susceptible nodes on its basic cycles with probability $\beta$.

In practice, the original SIR model differs from the cycle-based SIR model in that there is only a matrix, the former is based on the adjacency matrix while the latter is based on the spread matrix, as shown in Figure 1j. For node $i$, if node $j$ is on the basic cycles of node $i$, then $s_{ij} = 1$, otherwise $s_{ij} = 0$.

In addition, if we consider the effect of the direct edges in a basic cycle on the propagation probability, we can also define another spread matrix which considers both the cycles and stars structures. The spread matrix defined in the main text and the second spread matrix are denoted as $S^{\mathrm{I}}(G)$ and $S^{\mathrm{II}}(G)$, respectively. The $S^{\mathrm{II}}(G)$ is defined as follows.

**Spread matrix II**: let the ordinary network $G(V, E)$ and its spread matrix $S^{\mathrm{I}}(G)$. Then we modify all the elements $s_{ij}$ equal to 1 in $S^{\mathrm{I}}(G)$ according to the following rules, that is, if the element $a_{ij}$ of the same position in the adjacency matrix is equal to 0, then the element becomes a decimal. The calculation rule for this decimal is: first calculate the number of all indirect edges of each node $i$, that is, $m_i = \sum_j^N (c_{ij} - a_{ij})$, where $N$ is network size, $c_{ij}$ and $a_{ij}$ are the values of the corresponding position in the spread matrix $S^{\mathrm{I}}(G)$ and the adjacency matrix (see Figure 5a), respectively. Denote by $d_{\mathrm{norm}}$ as 1 plus the maximum value of $m_i$. Finally all decimals of the $i$-th row are equal to $m_i / d_{\mathrm{norm}}$, then $S^{\mathrm{II}}(G)$ is obtained.

In $S^{\mathrm{II}}(G)$, it is assumed that the propagation probability between the direct connections is higher than that between the indirect connections. Taking the network in Figure 1e in the main text as an example, its spread matrix II is shown in Figure 5b. Because edges (3, 7), (6, 8), (7, 3) and (8, 6) are indirect edges, so the numbers of indirect edges of the four nodes are all equal to 1, and $d_{\mathrm{norm}}$ is equal to 2.

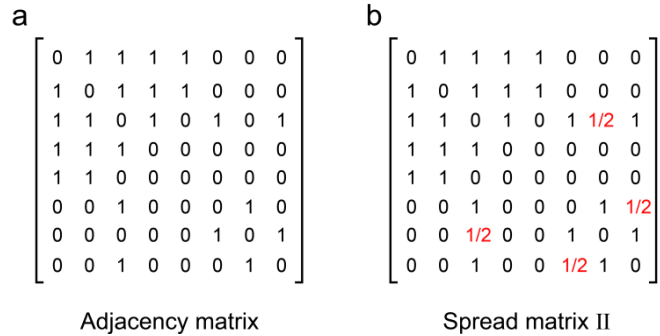

**Figure 5. The adjacency matrix and spread matrix II of the network in Figure 1e.** The red values are differences between the spread matrix I and the spread matrix II.

**Acknowledgments:** This work is supported by the National Natural Science Foundation of China (Nos. 11622538, 61673150), and the Zhejiang Provincial Natural Science Foundation of China (No. LR16A050001). L.L. acknowledge the Science Strength Promotion Programme of UESTC, Chengdu.

### Reference

1   Erdös, P. & Rényi, A. On random graphs. *Publ. Math. Debrecen* **6**, 290-297 (1959).

2   Erdős, P. & Rényi, A. On the evolution of random graphs. *Publ. Math. Inst. Hung. Acad. Sci*. **5**, 17-60 (1960).

3. Watts, D. J. & Strogatz, S. H. Collective dynamics of 'small-world' networks. *Nature* **393**, 440-442 (1998).

4. Barabási, A.-L. & Albert, R. Emergence of scaling in random networks. *Science* **286**, 509–512 (1999).

5. Shi, D., Chen, G., Thong, W. W. K. & Yan X. Searching for optimal network topology with best possible synchronizability. *IEEE Circ. Syst. Mag.* **13**, (1) 66-75 (2013).

6. Watson, B. Su Tung-p'o: Selections from a Sung dynasty poet. (Columbia Univ. Press, 1965).

7. Lü, L., Chen, D.-B. & Zhou, T. The small world yields the most effective information spreading. *New J. Phys.* **13**, 123005 (2011).

8. Sizemore, A. E. *et al*. Cliques and cavities in the human connectome. *J. Comput. Neurosci.* **44**(**1**) 115-145 (2018).

9. Berge, C. Graphs and hypergraphs. (North-Holland Pub. Co., 1973).

10. Lou, Y., Wang, L. & Chen, G. Toward Stronger Robustness of Network Controllability: A Snapback Network Model. *IEEE T. Circuits-I.* **99**, 1-9 (2018).

11. Bonneau, H., Hassid, A., Biham, O., Kühn, R. & Katzav, E. Distribution of shortest cycle lengths in random networks. *Phys. Rev. E* **96**(**6**), 062307 (2017).

12. Han, D. *et al.* Circular RNA MTO1 acts as the sponge of miR-9 to suppress hepatocellular carcinoma progression. *Hepatology* **66**, 1151 (2017).

13. Zhou, T., Lü, L. & Zhang Y.-C. Predicting missing links via local information. *Eur. Phys. J. B* **71**, 623–630 (2009)

14. Kitsak, M. *et al*. Identification of influential spreaders in complex networks. *Nat. Phys.* **6**, 888-893 (2011).

15. Lü, L., Zhou, T., Zhang, Q.-M. & Stanley, H. E. The H-index of a network node and its relation to degree and coreness. *Nat. Commun.* **7**, 10168 (2016).

16. Zomorodian, A., Carlsson, G. Computing persistent homology. *Discrete Comput. Geom.* **33**(**2**), 249-274 (2005).

17. Kendall, M. G. A new measure of rank correlation. *Biometrika* **30** (**1/2**), 81-93 (1938).

18. Achacoso, T. B. & Yamamoto, W. S. AY's Neuroanatomy of C. elegans for Computation. (CRC Press, 1992).

19. Guimerà, R., Danon, L., Díazguilera, A., Giralt, F. & Arenas, A. Self-similar community structure in a network of human interactions. *Phys. Rev. E* **68**, 065103 (2003).

20. Gutiérrez, J., Urbano, P. Accessibility in the European Union: the impact of the trans-European road network. *J. transp. Geogr.* **4** (**1**), 15-25(1996).

21. Newman, M. E. J. Finding community structure in networks using the eigenvectors of matrices. *Phys. Rev. E* **74**, 036104 (2006).

22. Jeong, H., Mason, S. P., Barabási, A.-L. & Oltvai, Z. N. Lethality and centrality in protein networks. *Nature* **411**, 41-42 (2001).

23 Hethcote, H. W. The mathematics of infectious diseases. *SIAM Rev.* **42**, 599-653 (2000).

24 Newman, M. E. J. Spread of epidemic disease on networks. *Phys. Rev. E* **66**(**1**), 16128 (2002).

25 Lü, L. & Zhou, T. Link prediction in complex networks: A survey. *Physica A* **390**, 1150-1170 (2011).

26 Pan, L., Zhou, T., Lü, L., & Hu, C.-K. Predicting missing links and identifying spurious links via likelihood analysis. *Sci. Rep.* **6**, 22955 (2016).

27 Shi, D., Lü, L. & Chen, G. Totally Homogeneous Networks. *Natl. Sci. Rev.* (to be published, 2019).

28 Castellano, C. & Pastor-Satorras, R. Thresholds for epidemic spreading in networks. *Phys. Rev. Lett.* **105**, 218701 (2010).

29 Shu, P., Wang, W., Tang, M. & Do, Y. Numerical identification of epidemic thresholds for susceptible-infected-recovered model on finite-size networks. *Chaos* **25**, 063104 (2015).

# SUPPLEMENTARY INFORMATION 1

**Algorithm for finding basic cycles**

Finding all basic 1-cycles (hereinafter referred to as cycle) in a network is time-consuming, especially in large-scale networks. If the entire network is traversed by nodes enumeration, the time and space costs are very huge. For this reason, inspired by the spanning tree or spanning forest of a given network, we design a fast cycle search algorithm, which greatly improves the search efficiency. The algorithm is mainly composed of the spanning tree construction process of network and the recursive backtracking process of the cycle search.

The spanning tree construction process: (i) for each node, there are five attributes: node number, degree, the insert time (namely number of adding order), the number of layers and the list of patriarchal nodes (the neighbors of the same layer or the previous layer are regarded as patriarchal nodes). At the beginning, randomly select a node as the root node (generally according to the node number), initialize its number of layers to 0, push -1 into its patriarchal nodes list, where -1 means that the node have no patriarchal node, and then access the root node (the layer where the current access node is called the current access layer), that is, all the neighbors of the root node are added to the tree according to the node number to form second layer (called the add-on layer) and the number of layer that are update to 1, and the root node is added to their lists of patriarchal nodes. Each time a node is accessed or added, the length of current access layer and the length of add-on layer (that are both equal to the number of nodes not accessed in their own layers) are also updated; (ii) when all the nodes in the current access layer have been accessed, the current access layer and its length are updated to add-on layer and its length respectively, the add-on layer and its length are updated to its own next layer (now empty) and 0, and then all nodes in the current access layer are accessed sequentially. Take node $i$ as an example: for all its neighboring nodes $j$, if node $j$ has been accessed, there is no operation; if the $j$ node is in the tree and is not accessed, then whether node $j$ is in the current access layer or in the add-on layer, a link will be added between $i$ and $j$. In addition, if node $i$ and $j$ are in the same layer, they are added to each other's patriarchal lists, otherwise only node $i$ is added to node $j$'s patriarchal list. Then record the pair of nodes ($i$, $j$) (or called the ring edge), because there must be at least one cycle passing through edge ($i$, $j$) in the existing tree. After all nodes in the current access layer have completed access, the backtracking search is started with these node pairs as the starting points; if node $j$ is not in the tree, add node $j$ to the add-on layer, initialize its patriarchal list and its number of layer, and update the corresponding layer length; (iii) repeat step (ii) to add and access other nodes in the connected component until all nodes in the current access layer have completed access and there are no more new nodes in the add-on layer at this time. So far, the construction process of the spanning tree is completed.

The recursive backtracking process of the cycle search calls different backtracking functions according to the positional relationship between the two nodes in the ring edge ($i$, $j$). If nodes $i$ and

*j* are on the same layer, the peer layer search function is called, otherwise, node *j* must be in the next layer of node *i* and the cross-layer search function is called. The specific backtracking process is (i) node *j* goes back one step to one of its patriarchal node *k*. If the *k* node is connected to *i* (or the times of backtracking are greater than 1 and nodes *i* and *j* are the same node), then a cycle (*i*, *j*, *k*) is found. Otherwise, the corresponding backtracking function is continuously selected according to the relationship between nodes *i* and *k*, and the backtracking operation is performed on the nodes with large insert time; (ii) every time a cycle is found, it is first judged whether it is a basic cycle, if not, it is discarded, instead the cycle is saved, meanwhile the upper semi-rings and lower semi-rings of all node pairs of non-directly connected in the same layer is found and recorded, both of them are simple paths with the two nodes as endpoints, but for the upper semi-rings, the layer of at least one node in this path is smaller than the layer of the two endpoints, for the lower semi-rings, the layer of any node is not less than the layer of the endpoints. When the next time back to the pair of nodes, the two types of semi-rings that have been recorded are respectively spliced with the currently obtained back-tracking path to get new basic cycles, and finally the back-tracking path obtained this time is recorded as a lower semi-ring of the pair of nodes; (iii) because the two types of semi-rings are stored, all the backtracking processes only need to back up the two layers from the current access layer, but if the two non-directly connected endpoints in the same layer have not been traced back, they should continue to backtrack to more upper layer.

Figure S1 gives an illustration of the search process on Figure 1e. Node 1 is set as the root node and first accessed, its four neighbors nodes 2, 3, 4 and 5 are sequentially added to the spanning tree, and append 1 to these leaf nodes' patriarchal node lists. The label on each edge represents their insert time. Node 2 will be accessed next, and labeled with 5 on the edge. At this time, because the neighbor node 3 of node 2 is already in the tree, at least one cycle is generated. Update patriarchal node lists of node 2 and 3, edge (2, 3) is recorded as a ring edge. Similarly, ring edges (2, 4) and (2, 5), and (3, 4) are also recorded when node 3 is assessed. When all nodes in the second layer have been accessed, the tree shown in Figure S1a is formed. Next, backtrack all ring edges recorded in this layer. For the ring edge (2, 3), node 3 traces back to node 1, as shown by the green edge in Figure S1b, since node 1 is in patriarchal nodes list of node 2, the a cycle {2, 3, 1} is found. Similarly, for edge (2, 4), node 4 goes back to 1 to get cycle {2, 4, 1}, or back to 3 to get {2, 4, 3}. For (2, 5), node 5 goes back to 1 to get {2, 5, 1}. For (3, 4), node 4 traces back to 1 to get the basic cycle {3, 4, 1}, or back to 2 to get {3, 4, 2} which is already obtained when considering edge (2, 3) and thus ignored as shown in Figure S1c. After all nodes in the third layer have been accessed, node 7 is added to the tree, for edge (8, 7) the node 7 traces back to the node 6, because there is no edge between node 6 and node 8, node 8 with larger insert time is selected to go back to node 3 and gets {8, 7, 6, 3}, as shown in Figure S1d.

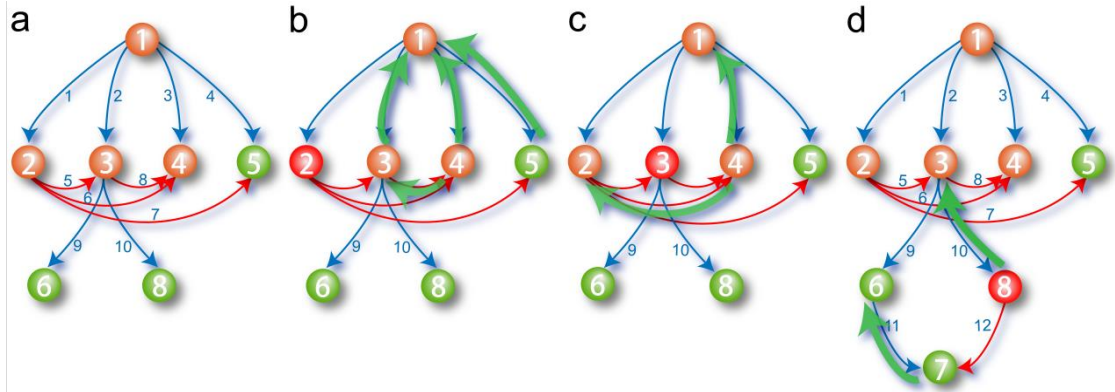

**Figure S1. The search process of basic cycles in Figure 1e.** Source nodes in backtracking process and ring edges are marked in red, and the backtracking edges are in green and bold.

For the smallest basic cycles of nodes, on this basis, we only need to stop the longer backtracking path in the backtracking process after the smallest cycles of all the nodes on one cycle have been found.

The algorithm constructs a tree in a breadth-first traversal manner and searches cycles in the depth-first traversal method to recursive backtrack, so the backtracking of each ring edge is asynchronous, and the backtracking branches (if any) generated in each step are also asynchronous. When one of the backtracking processes is completed, the other backtracking process does not need to repeat the invalid search process. Of course, this algorithm will have some local invalid search in the initial backtracking of the indirectly connected node pairs in the same layer, but this has little effect on the algorithm efficiency. This algorithm is very fast for small-scale sparse networks, but for larger-scale dense networks, the search time also increases due to the huge number of cycles in the networks.

# SUPPLEMENTARY INFORMATION 2

Here we simulate propagation experiments according to three different models, that is, the original SIR model, the cycle-based SIR model and the cycle-based SIR model considering the cycle length (denote as clSIR). The correlation coefficient τ between the three simulation results and the five indicators are calculated, see results in Table S1, Table S2 and Table S3. Note that the two cycle indicators are based on the node's smallest basic cycles in these three tables. In the clSIR model, we additionally consider the effect of the cycle length on the propagation probability. Assume that the smaller the basic cycle, the stronger the interaction between the nodes on the basic cycle. We use the reciprocal of the cycle length to express this effect. At this time, all elements equal to 1 in the spread matrix defined in the main text are multiplied by

$$z_i = \frac{1}{S_{max}} \cdot \sum_{m=1}^{C_{ij}} \frac{1}{L_m(i,j)},$$

where $C_{ij}$ is the number of co-cycle between the node $i$ and $j$, $L_m(i,j)$ is the length of the $m$-th co-cycle of nodes $i$ and $j$. $S_{max}$ is equal to the maximum of the term $S_i = \sum_{m=1}^{C_{ij}} \frac{1}{L_m(i,j)}$ for any node $i$ in the network, and the role of $1/S_{max}$ is to normalize $S_i$.

We simulate all β values in the range [0, 1] in increments of 0.01, and then selected the overall best groups as shown in Table S2 and S3. Three groups of results were compared and analyzed, we note that when the information disseminate according to star-structure in the original SIR model, the correlation coefficient between the three indicators based on the star-structure, such as *H*-index, and the simulation results is higher than others indices (see Table 1); when the information spread according to the cycle-structures, the correlation coefficient between the cycle number and the simulation result is higher than other indices (see Table S2 and S3). Based on the comprehensive analysis of the results, we find that, in terms of propagation of information or disease, the cycle number is more able to identify nodes with stronger propagation capabilities.

**Table S1 | The Kendall Tau between the ranking list given by each indicator and the corresponding node propagation capability given by the original SIR model.**

| Networks | Degree | *H*-index | Coreness | Cycle ratio | Cycle number |
|---|---|---|---|---|---|
| C. elegans (β = 0.09) | 0.8425 | **0.8554** | 0.8106 | 0.6096 | 0.7900 |
| Email (β = 0.14) | 0.8570 | **0.8827** | 0.8629 | 0.6017 | 0.7530 |
| EuroRoad (β = 0.50) | 0.4565 | 0.5164 | 0.5543 | 0.3706 | **0.5710** |
| NS-GC (β = 0.36) | 0.4650 | **0.4814** | 0.4461 | 0.1941 | 0.4594 |
| USAir (β = 0.06) | 0.8195 | **0.8497** | 0.8493 | 0.4303 | 0.8265 |
| Yeast (β = 0.15) | 0.7119 | **0.7577** | 0.7550 | 0.3918 | 0.6886 |

**Table S2 | The Kendall Tau between the ranking list given by each indicator and the corresponding node propagation capability given by the cycle-based SIR model.**

| Networks | Degree | H-index | Coreness | Cycle ratio | Cycle number |
|---|---|---|---|---|---|
| C. elegans ($\beta = 0.14$) | 0.8184 | 0.8299 | 0.7805 | 0.6540 | **0.8339** |
| Email ($\beta = 0.22$) | 0.7358 | 0.7296 | 0.7110 | 0.6995 | **0.8033** |
| EuroRoad ($\beta = 0.02$) | 0.5091 | 0.5719 | 0.7282 | **0.8319** | 0.7789 |
| NS-GC ($\beta = 0.06$) | 0.7168 | 0.7248 | 0.6851 | 0.4224 | **0.7269** |
| USAir ($\beta = 0.14$) | 0.8767 | **0.9075** | 0.9072 | 0.4951 | 0.8848 |
| Yeast ($\beta = 0.14$) | 0.7209 | 0.7173 | 0.7013 | 0.6058 | **0.7854** |

**Table S3 | The Kendall Tau between the ranking list given by each indicator and the corresponding node propagation capability given by the cycle-based and considering the cycle length SIR model.**

| Networks | Degree | H-index | Coreness | Cycle ratio | Cycle number |
|---|---|---|---|---|---|
| C. elegans ($\beta = 0.99$) | 0.7466 | 0.7761 | 0.7532 | 0.5921 | **0.8811** |
| Email ($\beta = 0.97$) | 0.7350 | 0.7549 | 0.7602 | 0.6442 | **0.8908** |
| EuroRoad ($\beta = 0.14$) | 0.5765 | 0.6246 | 0.7294 | 0.7276 | **0.9138** |
| NS-GC ($\beta = 0.22$) | 0.7747 | 0.8437 | 0.8219 | 0.4838 | **0.8500** |
| USAir ($\beta = 0.78$) | 0.8834 | 0.9075 | 0.9095 | 0.5146 | **0.9141** |
| Yeast ($\beta = 0.49$) | 0.7470 | 0.7707 | 0.7699 | 0.5201 | **0.8399** |

# SUPPLEMENTARY INFORMATION 3

Here we show the probability distribution of length of basic cycles in three real networks and three model networks. It can be seen that there is a significant difference between the real networks and the model networks.

Figure S2, Figure S3 and Figure S4 show the probability distributions of length of basic cycles in real networks, small-world model networks and scale-free model networks, respectively. Figure S2 shows that there are huge differences in the basic cycle distribution between different diverse real networks, and in general they are also different from the model networks. In Figure S3 the number of nodes is $N = 100$, and the number of edges is $E = 200$. When the rewiring probability $p$ is 0, the network is a regular ring lattice network, and when $p$ is 1, equivalent to a random network. In Figure S4 the number of nodes is $N = 100$ and the number of seed nodes is $m_0 = 5$, it can be seen that there are rich cycle-structures in the scale-free network, and as the $m$ increases, the number and variety of basic cycles increases rapidly.

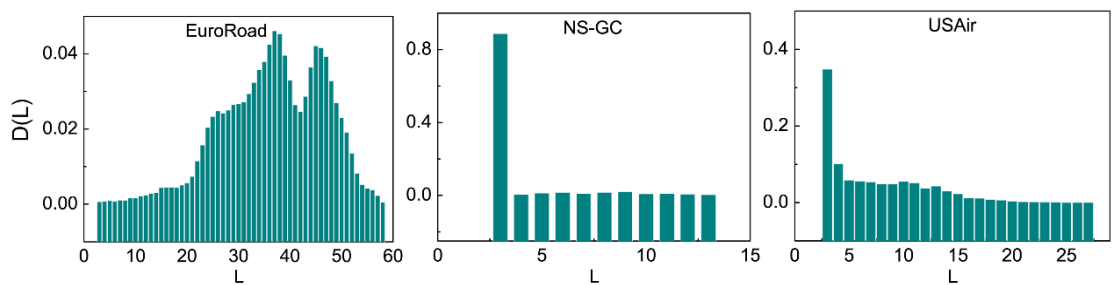

**Figure S2. Probability distribution of basic cycles with different length in real networks.** In the figures, the X axis represents length of basic cycles, the Y axis is its probability.

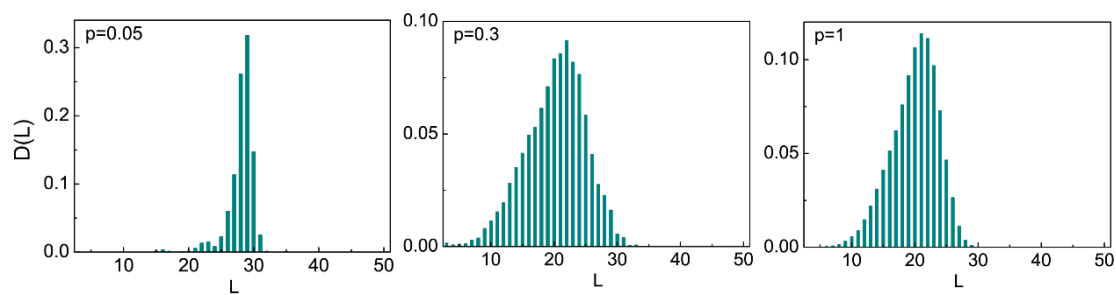

**Figure S3. Probability distribution of basic cycles' length with different rewiring probability $p$ in small-world networks.** In the figures, the X axis represents length of basic cycles, the Y axis is its probability.

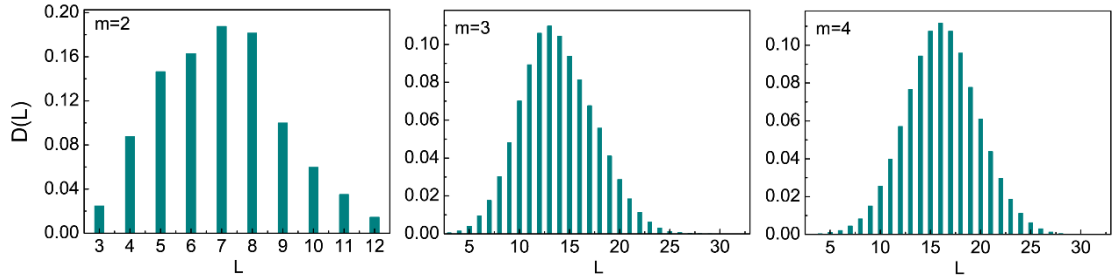

**Figure S4. Probability distribution of basic cycles' length with different number of adding edges *m* in scale-free model networks.** In the figures, the X axis represents length of basic cycles, the Y axis is its probability.

In the past two decades, the small-world model and the scale-free model have great significance for advancing the research of network science. In fact, these two important model networks are both related to cycles. For the small-world model network, in the evolution process from the regular network to the random network, the cycle changes also occur while the average shortest distance of the network becomes shorter. At the beginning most of them are fewer types of long-length cycles with different lengths, and by this time, the network has the better feature of small-worldness (for *p* = 0.05). With the increasing of random links, the types of cycles gradually increases, and the proportion of short-length cycles simultaneously increases. For the scale-free model network, when the number of adding edge in each time is *m* = 1, there is no cycle in the network, it is also a tree-like network which has only one shell according to network *K*-core decomposition; when $m \geq 2$, the rich cycle-structures are formed in the network, and as *m* increases, the distribution of the cycle lengths gradually becomes more uniform. Therefore, the evolution of the small-world network is a process of enriching the types of cycles and decreasing long-length cycles, the growth of the scale-free network is a process of forming various cycles and increasing long-length cycles.